\documentclass[aps,pra,superscriptaddress,reprint,showkeys,showpacs]{revtex4-1}
\usepackage{amsmath}    
\usepackage{graphicx}   
\usepackage{array}
\usepackage[dvipdfx,cmyk]{xcolor}     
\usepackage[caption=false]{subfig}  
\usepackage[colorlinks,linkcolor=red,citecolor=brown,urlcolor=blue]{hyperref}
\usepackage{dsfont,bm}

\newcommand{\be}{\begin{equation}}
\newcommand{\ee}{\end{equation}}
\newcommand{\ben}{\begin{eqnarray}}
\newcommand{\een}{\end{eqnarray}}
\newcommand{\bes}{\begin{subequations}}
\newcommand{\ees}{\end{subequations}}
\newcommand{\bF}{\begin{figure}}
\newcommand{\eF}{\end{figure}}

\newcommand{\ket}[1]{\vert{#1}\rangle}
\newcommand{\proj}[1]{\mbox{$|#1\rangle \!\langle #1 |$}}

\begin{document}

\title{A compact entanglement distillery using realistic quantum memories}

\author{Levon Chakhmakhchyan}
\affiliation{Laboratoire Interdisciplinaire Carnot de Bourgogne, UMR CNRS 6303 Universit\'{e} de Bourgogne, 21078 Dijon Cedex, France}
\affiliation{Institute for Physical Research, 0203 Ashtarak-2, Armenia}
\affiliation{A.I. Alikhanyan National Science Laboratory, Alikhanian Br. 2, 0036 Yerevan, Armenia}

\author{St\'{e}phane Gu\'{e}rin }
\affiliation{Laboratoire Interdisciplinaire Carnot de Bourgogne, UMR CNRS 6303 Universit\'{e} de Bourgogne, 21078 Dijon Cedex, France}

\author{Joshua Nunn}
\affiliation{Clarendon Laboratory, Department of Physics, University of Oxford, OX1 3PU, United Kingdom}

\author{Animesh Datta}
\affiliation{Clarendon Laboratory, Department of Physics, University of Oxford, OX1 3PU, United Kingdom}

\date{\today}

\begin{abstract}
We adopt the beam splitter model for losses to analyse the performance of a recent compact continuous-variable entanglement distillation protocol [Phys. Rev. Lett. \textbf{108}, 060502, (2012)] implemented using realistic quantum memories. We show that the decoherence undergone by a two-mode squeezed state while stored in a quantum memory can strongly modify the results of the preparatory step of the protocol. We find that the well-known method for locally increasing entanglement, phonon subtraction, may not result in entanglement gain when losses are taken into account. Thus, we investigate the critical number $m_c$ of phonon subtraction attempts from the matter modes of the quantum memory. If the initial state is not de-Gaussified within $m_c$ attempts, the protocol should be restarted to obtain any entanglement increase. Moreover, the condition $m_c>1$ implies an additional constraint on the subtraction beam splitter interaction transmissivity, \textit{viz.} it should be about $50\%$ for a wide range of protocol parameters. Additionally, we consider the average entanglement rate, which takes into account both the unavoidable probabilistic nature of the protocol and its possible failure as a result of a large number of unsuccessful subtraction attempts. We find that a higher value of the average entanglement can be achieved by increasing the subtraction beam splitter interaction transmissivity. We conclude that the compact distillation protocol with the practical constraints coming from realistic quantum memories allows a feasible experimental realization within existing technologies.
\end{abstract}


\maketitle

\section{Introduction}

Quantum entanglement, as one of the most intriguing features of quantum theory, remains a central area of research.
It is thought to be a valuable resource in quantum communication and information processing~\cite{review, review1, review2}
and is at the heart of protocols such as quantum teleportation~\cite{teleport, teleport1, teleport2}, and in quantum computation~\cite{comp}
and quantum cryptography~\cite{prot, prot1}. The key requirement in essentially all practical applications of quantum information science is
to protect the coherence of quantum states against decoherence, induced by the uncontrolled influences of an environment. One of the possible
strategies to protect entanglement from decoherence effects is based on reducing the interaction with the environment, which however, is not
sufficient in a number of cases. A more powerful method, which allows one to distribute entanglement between distant parties, involves the notion
of extracting a small ensemble of more strongly entangled states from a larger ensemble of weakly entangled states. The strategy achieving this is
referred to as {\it entanglement distillation}~\cite{dist1, dist2}. Challenging experimental realizations of these proposed distillation protocols
in qubit systems and other finite-dimensional settings \cite{finite,fdist} have been performed.

An important step in the theory of entanglement distillation was the invention of continuous-variable entanglement distillation protocols~\cite{cont, cont1,earl}. The essential entanglement resource here is provided through the Gaussian two-mode squeezed state, which is relatively easy to produce in the laboratory. These states are widely used in the realization of several other quantum protocols, including dense coding, entanglement swapping and others \cite{squeez, squeez1, squeez2}. The continuous-variable entanglement distillation protocols~\cite{cont, cont1} circumvent restrictions imposed by the no-go theorem relating to the distillation of entanglement using only Gaussian local operations and classical communication \cite{nogo, nogo1, nogo2}. Namely, the proposed procedure offers a complete distillation of Gaussian states to (asymptotically exact) Gaussian states, but via non-Gaussian territory. Implementation of some elements of the scheme has already been reported~\cite{elem, elem1, elem2, elem3}, though a full demonstration is still lacking. This is due to demanding technical specifications \cite{req} and the exponential resource requirements of the protocol. The ``de-Gaussification" of the initial state plays a crucial role in this protocol and can be achieved by subtracting a photon from the initial state. Note that this process itself can increase the  entanglement in the state. Different strategies for enhancing quantum entanglement in continuous-variable systems within this technique have been found~\cite{strat,earl}.

A continuous-variable entanglement distillation protocol using quantum memories was recently proposed by some of us~\cite{dist}. The scheme achieves the same increase in entanglement using much fewer resources than required by the earlier procedures \cite{cont, cont1}. This compact protocol involves only four quantum memories, allowing one to store results from previous iterations while the subsequent ones succeed~\cite{memories}, bringing about an exponential advantage in the run time. Additionally, one can repeatedly perform probabilistic operations on the same copy of a quantum state, without starting the whole process anew as in the case of a failure of a probabilistic local operation. Another resource-saving comes from the use of an asymmetric ``pumping" strategy: in contrast with earlier schemes, where two identical states are combined at each stage, in the compact protocol one iteratively operates on a state with a fixed resource state. This approach avoids the requirement of successfully distilling two copies of the $i^\mathrm{th}$ iterate at each stage, while still producing states with the same degree of entanglement at each iteration. Although a comprehensive comparison is challenging, the use of this pumping approach provides a near-exponential saving in the spatial resources required to reach a given degree of entanglement. As mentioned above, the compact protocol requires constant resources: just four memories.

The compact protocol as described here and in Ref.~\cite{dist} results in non-Gaussian states, even in the limit of a large number of iterations. Intuitively this is because the non-Gaussian resource state is repeatedly admixed with the current iterate in the same proportions. Interestingly, a modified version of the protocol that asymptotically produces Gaussian states, identical with the output of \cite{cont, cont1}, was proposed and analysed in Ref.~\cite{earl}. Here the Gaussian convergence is derived by means of a generalised central limit theorem for quantum operations. At each stage of the protocol, the degree of mixing between the current iterate and the resource state is reduced so as to allow the output to ``relax" to a Gaussian state. Although we have not studied this modified ``pumping Gaussifier'' protocol here, we expect the effects of realistic losses reported here to apply similarly to the protocol proposed in Ref.~\cite{earl}.

Generally, once locally prepared and stored in a quantum memory, two-mode squeezed states or any quantum states are subject to decoherence and deteriorate, with a consequent loss of entanglement. Transmission losses in all-optical continuous-variable entanglement distillation protocols have been studied previously \cite{lund}. In the present paper we address the question of how a realistic quantum memory affects the performance of the compact continuous-variable entanglement distillation protocol proposed in \cite{dist}. Note that the decoherence dynamics (loss) we are interested in does not alter the Gaussian character of the state \cite{loss}, and can be considered as a Gaussian channel. The approach that we use here for incorporating the dissipation dynamics is based on the so-called beam splitter model for losses. The latter considers a quantum memory as a set of beam splitters, with a lost reflected mode. The model is applicable in the Markovian limit, when the system dynamics is much slower than the one of the environment. Note that the dissipation drives the initial pure two-mode squeezed state to a mixed one. For quantifying entanglement of such states we use the logarithmic negativity \cite{negat, negat1}.

In this paper, we show that the dissipation can modify strongly the preparatory ``malting" step of the protocol~\cite{dist}, which consists in de-Gaussifying the initial two-mode squeezed state (distributed over two quantum memories), by retrieving a single photon from the quantum state stored in a memory. The spin-wave-type excitations of the quantum memories we are considering are similar to optical phonons in condensed matter systems, and so we refer to this operation as ``phonon subtraction". Particularly, we point out that one may not find entanglement increase after de-Gaussifying the initial state, which is not the case in a perfect memory. This effect puts additional constraints on the memory time-bandwidth product and the subtraction probability, which, however, do not complicate much the experimental realization of the protocol. We also show that high values of the average entanglement of the output state can be achieved by increasing the subtraction strength.

The paper is organized as follows: in Sec.~\ref{model} we introduce the beam splitter model for losses and detail its application on the storage of a two-mode squeezed state (TMSS) in quantum memories. In Sec.~\ref{malt} we incorporate the phonon subtraction from the TMSS and analyze the effects of decoherence on the preparatory malting step of the distillery. The full compact continuous-variable entanglement distillation protocol in the presence of dissipation, and some constraints on its experimental realization, are discussed in Sec.~\ref{full}. We conclude in Sec.~\ref{conc}.

\section{A beam splitter model for losses in a quantum memory}\label{model}

\begin{figure}[h]
\begin{center}
\includegraphics[width=7cm]{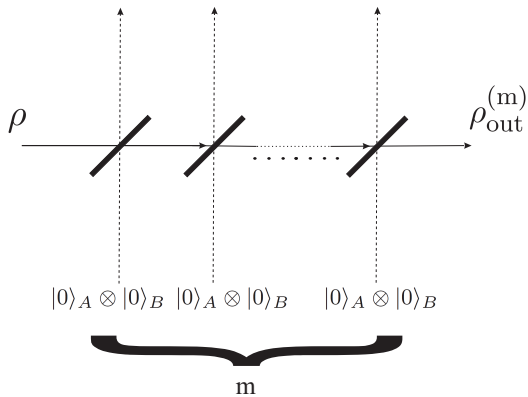}
\caption {The beam splitter model for losses (schematically). Decoherence within a quantum memory is modeled as a set of $m$ beam splitters with a state $\rho$ and a two-mode vacuum $|0\rangle_A\otimes|0\rangle_B$ as inputs. \label{beam}}
\end{center}
\end{figure}

Consider an arbitrary two-mode density matrix, represented in the number basis $|n\rangle$ by coefficients $p_{n, m, k, l}$ (subscripts $A$ and $B$ differentiate the modes):
\begin{equation}
\rho=\sum_{n, m, k, l=0}^\infty p_{n, m, k,l} |n\rangle_A \otimes |m\rangle_B \langle k|_A  \otimes \langle l|_B. \label{1}
\end{equation}
We are interested in modeling losses of such a state, when the latter is distributed over two quantum memories. For that we use a beam splitter model for losses, i.e., we consider the quantum memory as a set of beam splitters, with the state $\rho$ and a two-mode vacuum $|0\rangle_A\otimes|0\rangle_B$ as inputs (Fig.~\ref{beam}). In the present model the reflected mode accounts for the loss.

A beam splitter has two input modes $a_1$ and $a_2$, and implements the transformation
\begin{eqnarray}
\left(
  \begin{array}{c}
    a_1' \\
    a_2' \\
  \end{array}
\right)=U \left(
  \begin{array}{c}
    a_1 \\
    a_2 \\
  \end{array}
\right)\label{2}
\end{eqnarray}
with
\begin{eqnarray}
U=\left(
    \begin{array}{cc}
      t & r \\
      -r & t \\
    \end{array}
  \right),\label{3}
\end{eqnarray}
where $t^2$ and $r^2$ are, respectively, the transmissivity and reflectivity: $t^2+r^2=1$ (hereafter we refer to $t$ and $r$ as simply the transmissivity and reflectivity). Thus, mixing a two-mode vacuum $|0\rangle_A\otimes|0\rangle_B$ and the state (\ref{1}) on a beam splitter, one obtains the output state
\begin{widetext}
\begin{equation}
\rho_{\mathrm{out}}'=\!\!\!\!\!\!\!\!\sum_{n,m,k,l=0}^\infty \sum_{K=0}^{n,m,k,l}  p_{n, m, k, l} A_{n,k_1}(t) A_{m,k_2}(t) A_{k,k_3}(t) A_{l,k_4}(t)|n-k_1, k_1\rangle_A |m-k_2, k_2\rangle_B \langle k-k_3, k_3|_A \langle l-k_4, k_4|_B, \label{4}
\end{equation}
\end{widetext}
with $K=\{k_1,k_2,k_3,k_4\}$ representing the indices,
\begin{eqnarray}
&A_{qq'}(t)=\sqrt{\left(\begin{array}{c}
        q \\
        q'
      \end{array}\right)}t^{q-q'} r^{q'}
\end{eqnarray}
and $\left(\begin{array}{c}
        q \\
        q'
      \end{array}\right)$ are the binomial coefficients. The next step in the beam splitter model for losses is the trace out operation over the reflected (lost) mode, which yields
\begin{widetext}
\begin{equation} \label{5}
\rho_{\mathrm{out}}^{(1)}=\sum_{n, m, k, l=0}^\infty \sum_{k_1=0}^{\mathrm{min}(n, k)} \sum_{k_2=0}^{\mathrm{min}(m, l)} p_{n, m, k, l} A_{n,k_1}(t) A_{m,k_2}(t) A_{k,k_1}(t) A_{l,k_2}(t) |n-k_1\rangle_A |m-k_2\rangle_B \langle k-k_1|_A \langle l-k_2|_B.
\end{equation}
\end{widetext}
Hereafter, the superscript of the output state labels the number of loss events (beam splitters) that the initial state has undergone. Note that photons may or may not be lost at each loss event, but purity of the state decreases after each loss event. The general expression (\ref{5}) can be used iteratively, considering the output state of the $i^{\mathrm{th}}$ beam splitter as an input for the $(i+1)^{\mathrm{th}}$ one. Note, that the transmissivity $t$ can be also related to the time-bandwidth product $\tau$ of a quantum memory: $\tau=1/(1-t^2)$. Below we use the latter quantity, since it is a more convenient parameter for characterizing a memory: it determines the number of iterations that can be executed within the coherence lifetime of the memory.

In the present work, we are mainly interested in a two-mode squeezed state $\rho_{\mathrm{TMSS}}=\proj{\Psi_{\mathrm{TMSS}}}$
\be
\ket{\Psi_{\mathrm{TMSS}}}=\sqrt{1-\lambda^2}\sum_{n=0}^\infty\lambda^{n}\ket{n}_A\ket{n}_B,
\label{6}
\ee
where $\lambda$ is the squeezing parameter (for the compact distillery protocol~\cite{dist}, $|n\rangle$ corresponds to the number of phonons in the matter mode). Using Eq.~(\ref{5}), after one loss event for this particular input state we obtain
\begin{widetext}
\be
\rho_{\mathrm{out}}^{(1)}=(1-\lambda^2)\sum_{n, n'=0}^\infty\lambda^{n+n'} \sum_{k,k'=0}^{\mathcal{N}} A_{n,k}(t) A_{n,k'}(t) A_{n',k}(t) A_{n',k'}(t)|n-k\rangle_A |n-k'\rangle_B \langle n'-k|_A \langle n'-k'|_B,
\label{7}
\ee
\end{widetext}
where $\mathcal{N}=\mathrm{min}(n, n').$ After $m$ loss events, a two mode squeezed state is mapped into
\begin{widetext}
\begin{eqnarray}\nonumber
\rho_{\mathrm{out}}^{\mathrm{(m)}}&&=(1-\lambda^2)\sum_{n, n'=0}^{\infty}\lambda^{n+n'}\sum_{k_1, k_1'=0}^{\mathrm{min}(n, n')}\sum_{k_2, k_2'=0}^{\mathrm{min}(n-k_1, n'-k_1)}...\sum_{k_m, k_m'=0}^{\mathrm{min}(n-k_1-k_2-...-k_{m-1}, n'-k_1-k_2-...-k_{m-1})}\\ \nonumber
&&A_{n,k_1}(t) A_{n,k_1'}(t) A_{n',k_1}(t) A_{n',k_1'}(t) A_{n-k_1, k_2}(t) A_{n-k_1', k_2'}(t)A_{n'-k_1, k_2}(t) A_{n'-k_1', k_2'}(t)\\
&& \times~A_{n-k_1-k_2-...-k_{m-1}, k_m}(t)A_{n-k_1'-k_2'-...-k_{m-1}', k_m'}(t)A_{n'-k_1-k_2-...-k_{m-1}, k_m}(t)A_{n'-k_1'-k_2'-...-k_{m-1}', k_m'}(t)  \\ \nonumber
&&\times |n-k_1-...-k_m\rangle_A |n-k_1'-...-k_m'\rangle_B~\langle n'-k_1-...-k_m|_A \langle n'-k_1'-...-k_m'|_B. \label{8}
\end{eqnarray}
\end{widetext}
Although, this formula is analytical, it is not convenient for quantitative analyzes. Thus in what follows, we use the recursive approach, based on Eq.~(\ref{5}).

\section{Malting step with losses}\label{malt}

In this section we show how the losses affect the preparatory (malting) step of the original compact distillery protocol, described in Ref.~\cite{dist}. The malting procedure results in a non-Gaussian output state, prepared from an initial two-mode squeezed state (\ref{6}). Namely, the latter undergoes a non-Gaussian operation of a phonon subtraction from each of the modes, which can be achieved by sending in weak control pulses and detecting the emission of a photon at the output. The quantum memories we are considering are based on Raman interactions in atomic ensembles, where light can be stored into a material excitation (phonon mode) and retrieved from it on-demand, by the application of ancillary ``control pulses" that drive the Raman scattering. There is a formal correspondence between the Raman interaction, which couples optical and material modes, and a standard beam splitter interaction between two optical modes~\cite{mem}. Making use of this analogy, we are able to implement phonon subtraction by applying a weak control pulse to the memory, which partially retrieves the stored excitation onto a photon detector. This is equivalent to placing a highly transmissive beam splitter in the path of an all-optical state, and reflecting a small portion of it towards a detector, which is the standard procedure for photon subtraction~\cite{elem1}.

\begin{figure}[h!]
\begin{center}
\includegraphics[width=9cm]{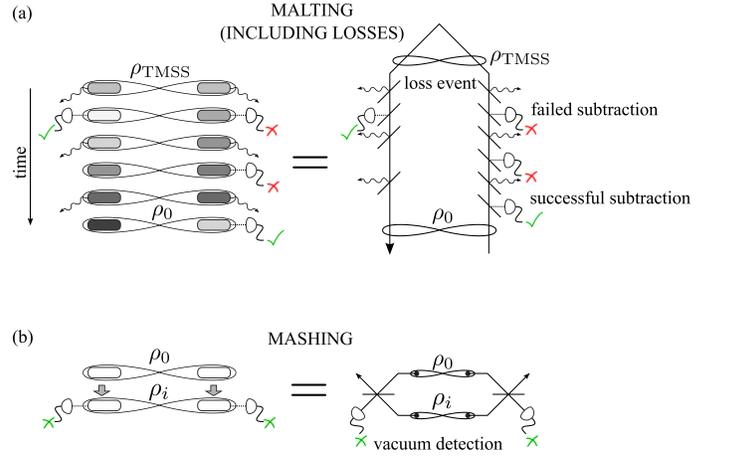}
\caption {(a) Malting step with losses. The initial two-mode squeezed state $\rho_{\mathrm{TMSS}}$ undergoes phonon subtraction attempts (achieved by sending in weak control pulses, which implement the beam splitter interaction, and by detecting the emission of a photon at the output), with interstitial loss events (modeled by a beam splitter model for losses). The state is considered to be ready (malted), when phonons have been retrieved from both modes. The left hand side shows the scheme with memories, and the right hand side shows the all-optical analog. (b) Mashing step. At the $i^{\mathrm{th}}$ iteration of the mashing step a state $\rho_0$ is mapped into the memories containing the state $\rho_i$, with vacuum detected in the transmitted modes. This is equivalent to the all-optical scheme shown on the right, where the states are interfered on beam splitters with subsequent vacuum detection. \label{diagram}}
\end{center}
\end{figure}

Here we assume that a discrete loss event occurs between two consecutive phonon subtraction attempts. More precisely, malting starts with one loss event, followed by an attempt of a phonon subtraction on both arms. If the attempt was not successful on either of the modes (i.e. if a vacuum detection took place on both arms), another loss event happens, followed by a consecutive phonon subtraction attempt, and so on. When one detects a phonon at one of the arms, it is left to undergo subsequent loss events, with loss and vacuum detection taking place on the other mode, until a successful phonon retrieval [Fig.~\ref{diagram}(a)]. The non-Gaussian output state is considered to be ready when phonons have been subtracted from both modes.

If a given state (\ref{1}) undergoes a detection of $q$ phonons by an ideal number-resolving detector, the unnormalized output state reads
\ben
\rho_{\mathrm{sub}}'&=&\sum_{n, m, k, l=0}^\infty A_{n, q}(t_s) A_{m, q}(t_s) A_{k, q}(t_s) A_{l, q}(t_s) \nonumber\\
 && \times |n-q\rangle_A |m-q\rangle_B \langle k-q|_A \langle l-q|_B, \label{9}
\een
where $t_s$ stands for the transmissivity of the subtraction beam splitter interaction. The trace of $\rho_{\mathrm{sub}}'$ gives the probability of a detection of $q$ phonons, and the normalized output state reads $\rho_{\mathrm{sub}}=\rho_{\mathrm{sub}}'/\mathrm{Tr}\left(\rho_{\mathrm{sub}}'\right)$. In the present protocol we deal only with vacuum detection $(q=0)$ and a single phonon subtraction $(q=1)$.

For quantifying entanglement we use the logarithmic negativity, defined as
\begin{equation}
N(\rho)=\log_2\left|\left|\rho^{T_A}\right|\right|, \label{10}
\end{equation}
where $\left|\left|\rho^{T_A}\right|\right|$ is the trace norm of the partial transposed $\rho^{T_A}$ of a bipartite density matrix $\rho$ \cite{negat, negat1}.

\begin{figure}[h!]
\begin{center}
\includegraphics[width=9cm]{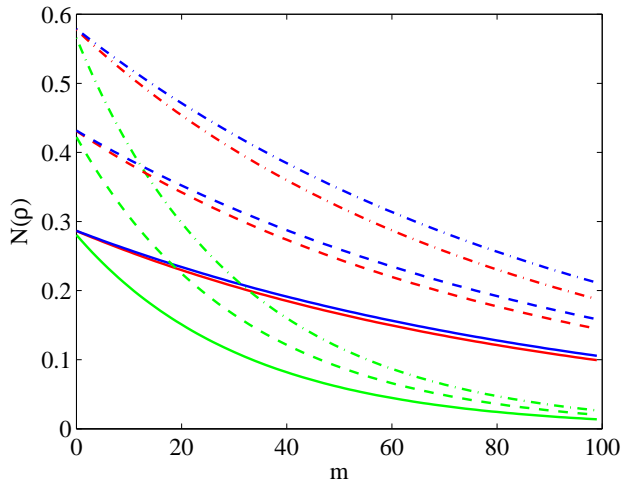}
\caption {Logarithmic negativity $N(\rho)$ of a two-mode squeezed state (\ref{6}), versus discrete time step (memory clock-cycle) $m$, for $\lambda=0.1$ (full curves), $\lambda=0.15$ (dashed curves) and $\lambda=0.2$ (dotted-dashed curves). Blue curves: the state undergoes $m$ vacuum detections with $\tau=1/(1-t^2)=100$, $t_s=t$; red curves: the state undergoes $m$ losses with $\tau=1/(1-t^2)=100$, $t_s=t$; green  curves: the state undergoes $m$ losses and vacuum detections with $\tau=1/(1-t^2)=100$, $t_s=0.99$. \label{compare}}
\end{center}
\end{figure}

In Fig.~\ref{compare} we show how the entanglement [quantified by $N(\rho)$] of a two-mode squeezed state (\ref{6}) evolves in a discrete time scale if either vacuum detection or loss or both occur. The curves start from a point corresponding to the amount of entanglement of a two-mode squeezed state: $N(\rho)=\log_2\frac{1+\lambda}{1-\lambda}$. As expected, in all of these three cases the entanglement decreases exponentially versus time. However, for equal values of $t$ and $t_s$ the loss affects the rate of decrease more than the vacuum detection. When combined together, these two effects bring about a steeper slope.

\begin{figure}[h!]
\begin{center}
\includegraphics[width=9cm]{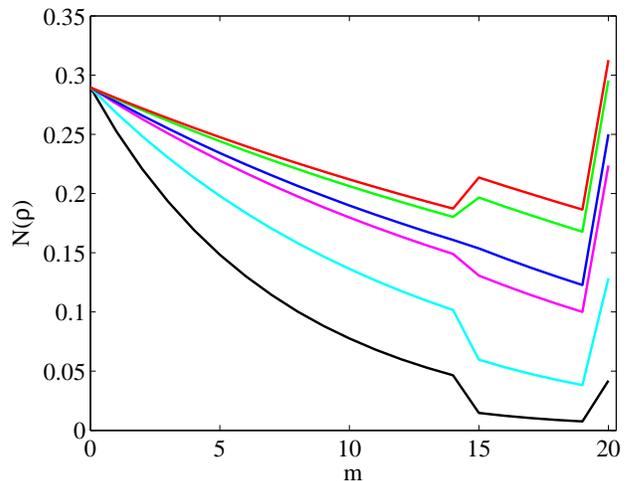}
\caption {Logarithmic negativity $N(\rho)$ of a two-mode squeezed state (\ref{6}), versus discrete time step (memory clock-cycle) $m$, for $\lambda=0.1$, $t_s=0.99$ and different values of $\tau=1/(1-t^2)$: $\tau=100, 80, 50, 40, 20, 10$ from the upper to the lower curve. The phonon subtraction occurs at $m_A=15$ on the mode $A$, and at $m_B=20$ on the mode $B$. \label{one}}
\end{center}
\end{figure}

Note that in an ideal memory one finds an increase of entanglement even after a phonon subtraction on only one arm. However, as we show below, this is not the case if losses are taken into account. Figure~\ref{one} shows the evolution of entanglement versus the memory clock-cycle $m$. Although here we suppose that the phonon subtraction happens at a fixed step on each arm ($m_A=15$, and $m_B=20$), the picture remains qualitatively the same for any values of $m_{A}$ and $m_{B}$.

Additionally, as one finds here, the smaller $\tau$ is, the less benefit of entanglement increase we get. In other words, the decoherence damages the initial state, driving its properties away from that of a two-mode squeezed state. Furthermore, for extremely low values of the time-bandwidth product, one does not achieve any entanglement increase through a successful malting step. In this case the distillation attempt can be considered to be ineffective. In the next section we return to the question of the possible number of phonon subtraction attempts.

\begin{figure}[h!]
\begin{center}
\includegraphics[width=10cm]{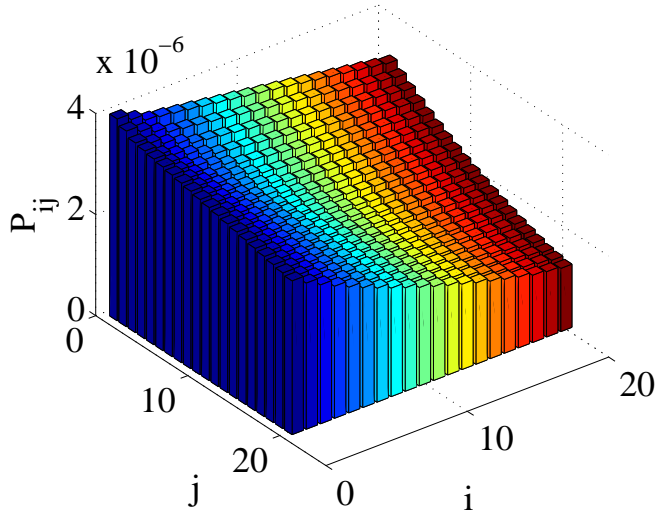}
\caption {The matrix $P_{ij}$ of probabilities of a phonon subtraction at $i^\mathrm{th}$ and $j^\mathrm{th}$ step on the first and second arms respectively. Here $\lambda=0.1$, $\tau=1/(1-t^2)=100$ and $t_s=0.99$ ($i, j=1, ..., 20$). \label{prob}}
\end{center}
\end{figure}

Obviously, the phonon subtraction is a probabilistic process. Thus, it is of interest to study the behavior of the probability $P_{ij}$ of a phonon subtraction, depending on $i$ and $j$, i.e., on the index number, at which a detection occurs on the arms $A$ and $B$ respectively. The success probability is given by the trace of the matrix $\rho_{\mathrm{sub}}'$ in Eq.~(\ref{9}), with $q=1$. In Fig.~\ref{prob} we show the matrix $P_{ij}$. It is symmetric, which confirms the symmetry of the distillation protocol with respect to the modes $A$ and $B$. Secondly, the probability $P_{ij}$ decreases exponentially with $i$ and $j$. Thus, most probably the two-mode squeezed state is malted at the first attempt, which additionally guarantees the least amount of entanglement loss due to decoherence.

\section{Mashing step and the full distillation protocol}\label{full}

After the initial two-mode squeezed state has been malted in two pairs of quantum memories, one proceeds to the iterative mashing step. The resource state obtained through malting is denoted here by $\rho_0$. Note that the Fock decomposition of the state $\rho_0$ retains its original Fock space structure in the ideal distillery protocol. However, when losses are taken into account, the Fock decomposition of $\rho_0$ is not similar to that of the two-mode squeezed state: the decoherence damages its original structure.

In brief, the mashing step involves the following operations. In the first step of its iteration, two copies of the state $\rho_0$ are combined on two 50/50 beam splitters. In the case that each party detects vacuum on one of the emerging modes from each beam splitter, the resultant state in the other two modes is $\rho_1$. Next, $\rho_1$ is interfered with a fresh copy of $\rho_0$ to produce $\rho_2$ upon a vacuum detection, and so on. At stage $i$ of the protocol, we combine $\rho_i$ with $\rho_0$ on beam splitters and detect vacuum, to produce the state $\rho_{i+1}$ [Fig~\ref{diagram}(b)]. This iterative procedure can be expressed as
\begin{eqnarray}
\rho_{i+1}=\langle \mathrm{vac} |W\rho_i^{A_2 B_2} \otimes \rho_0^{A_1B_1}W^\dagger|\mathrm{vac} \rangle,
\end{eqnarray}
where $|\mathrm{vac}\rangle = |0_{A_1}\rangle \otimes |0_{A_2}\rangle$ is the joint state produced by vacuum detection, and where $W = U_{A_1 A_2}\otimes U_{B_1 B_2}$, with $A_1$, $A_2$, $B_1$ and $B_2$ denoting the four memories used in the protocol and the matrices $U$ corresponding to 50/50 beam splitters.

In the above described scheme of the mashing step we do not take into account the decoherence affecting a state $\rho_i$, while it awaits new states to malt. Although being an approximation, this assumption keeps the calculations simple, still providing a good insight into the effects of losses of the protocol. Additionally, as shown below, a large part of the possible entanglement increase comes from the malting step. Thus one can consider the above simplification as a good starting point aimed towards the understanding of loss effects in the present distillation protocol.

\begin{figure}[h!]
\begin{center}
\includegraphics[width=9cm]{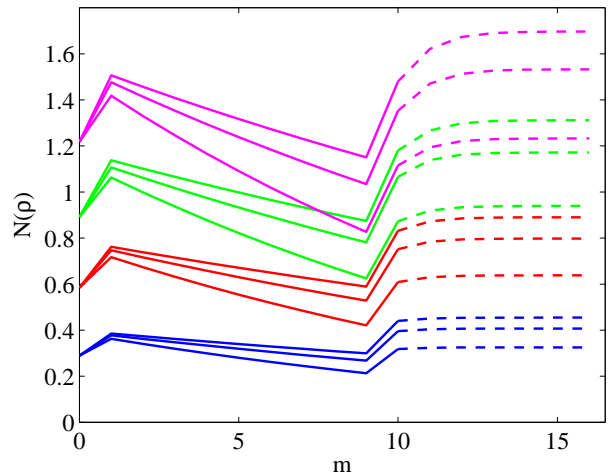}
\caption {The negativity $N(\rho)$ versus the full distillation protocol clock-cycle number $m$ for $\lambda=0.1$ (blue curves), $\lambda=0.2$ (red curves), $\lambda=0.3$ (green curves) and $\lambda=0.4$ (magenta curves). Here $\tau=1/(1-t^2)=100$ and $t_s=0.99, 0.98, 0.96$ from the upper to the lower curve of each block of the same color. The phonon subtraction occurs at $m_A=1$ and $m_B=10$. The full curves correspond to the malting step, while the dashed ones stand for mashing. \label{dist}}
\end{center}
\end{figure}

In Fig.~\ref{dist} we show how the full distillation protocol modifies the entanglement properties of an initial two-mode squeezed state. Firstly, we find that the above mapping, transforming $\rho_{i}$ to $\rho_{i+1}$, converges to its fixed point after quite a small number of iterations. Secondly, the mashing step does not increase the entanglement of a malted state by very much for low values of the squeezing parameter. Meanwhile, its impact becomes significant for relatively high values of $\lambda$.

In Fig.~\ref{dist} we have fixed the number of attempts after which a successful phonon subtraction occurs. However, as already mentioned in the previous section, a long storage in a lossy memory during the malting step can result in an absence of a positive entanglement gain. For moderate amount of losses in a memory, this corresponds to a relatively large number of unsuccessful subtraction attempts. It is of interest to consider the latter's threshold number, after which the protocol should be restarted. More precisely, we are interested in the critical value $m_c$ of subtraction attempts, within which we must succeed if we are to have a gain in entanglement in the full distillation protocol. Here we propose two possible definitions of this quantity:

\begin{enumerate}

\item $m_c$ is the critical number of simultaneous detection attempts on both arms

\item $m_c$ is the critical number of subtraction attempts on one of the arms, when the index $m_i$ ($i=A, B$) of a successful attempt on the opposite arm is considered to be fixed.

\end{enumerate}

Below we study the second option. Additionally, we assume that a successful phonon subtraction on one of the arms occurs at the first clock-cycle. The reason for this choice is that it leads to a maximal entanglement gain and has the highest probability. Note that this option is still symmetric, in a sense that the value of $m_c$ does not depend on whether it is obtained at a fixed $m_A$ or $m_B$.

\begin{figure}[ht]
\begin{center}
\small{(a)} \includegraphics[width=0.45\textwidth]{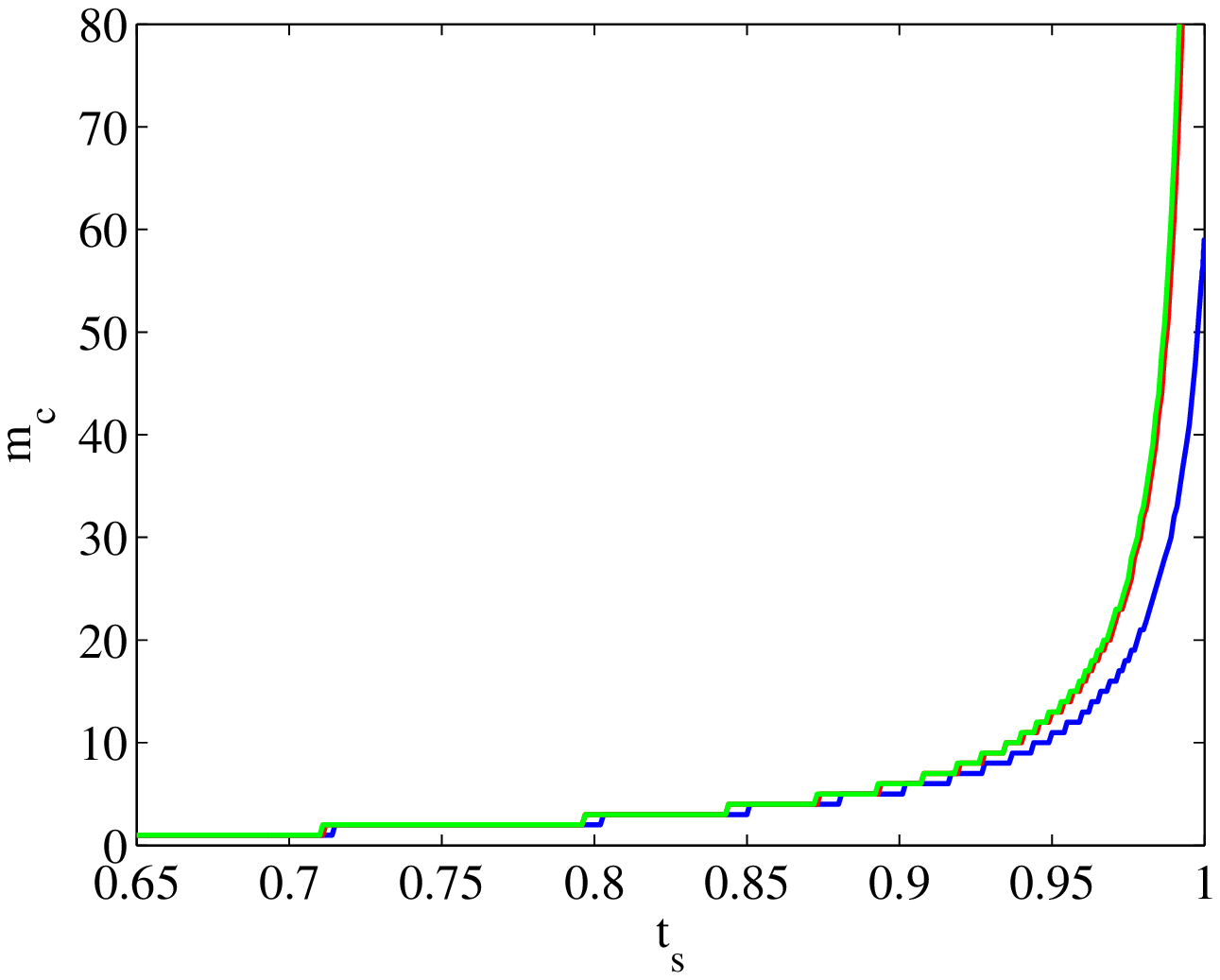}\\
\small{(b)} \includegraphics[width=0.45\textwidth]{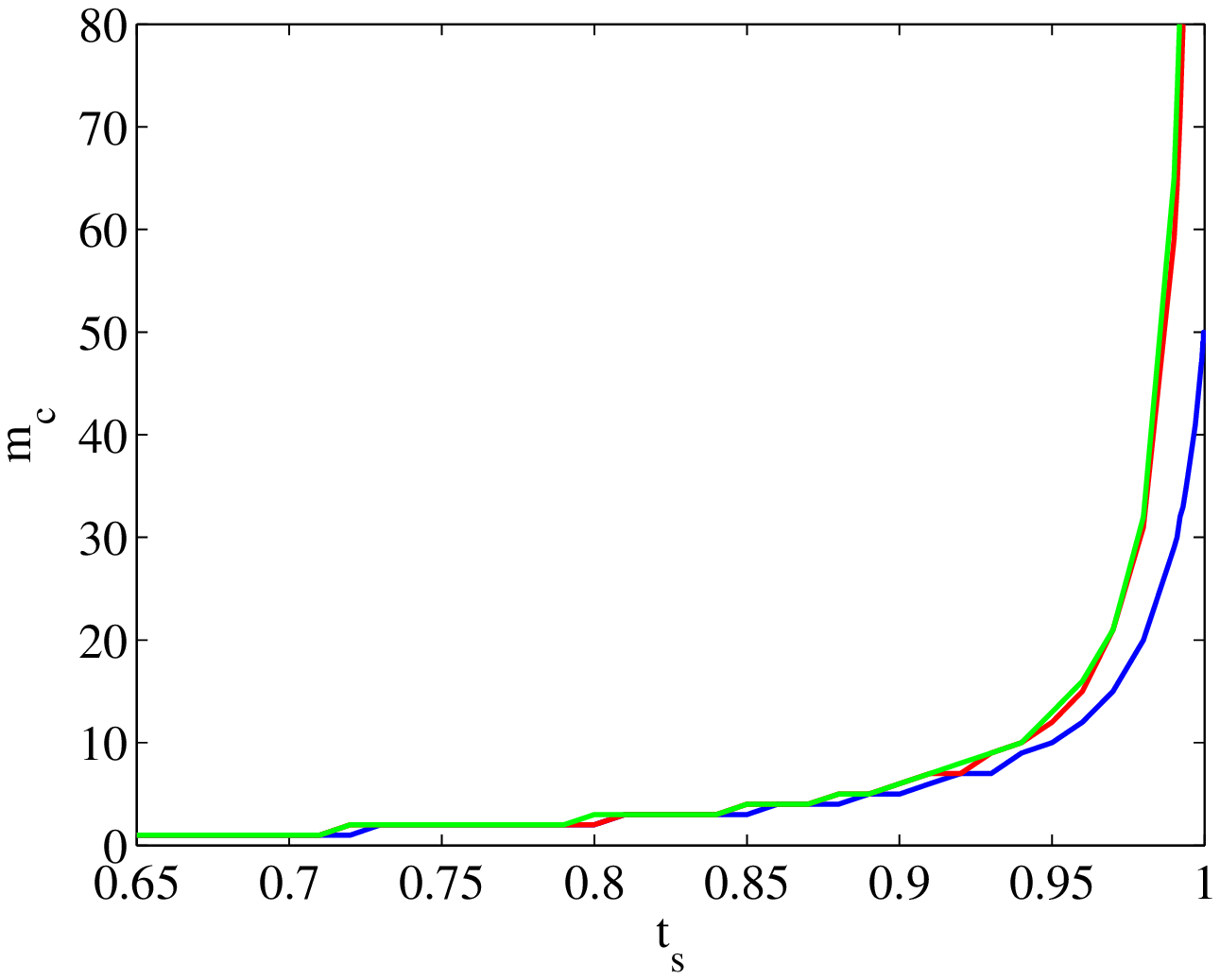}
\caption {The critical number of phonon subtraction attempts $m_c$, versus transmissivity $t_s$ for (a) $\lambda=0.1$ and (b) $\lambda=0.2$. A successful subtraction on the first arm is considered to occur at the first clock-cycle. Here  $\tau=1/(1-t^2)=100$ (blue curve), 1000 (red curve) and 10000 (green curve). \label{mc}}
\end{center}
\end{figure}

Figure~\ref{mc} shows how the above defined quantity $m_c$ depends on the transmissivity $t_s$.
We find that there exists a critical value $t_s^c$ such, that if $t_s<t_s^c$, one does not have any increase of entanglement through the distillation protocol, even if the state was malted at the very first phonon subtraction attempt. Our calculations show that $t_s^c\approx0.7$ for a wide range of parameters $\tau$ and $\lambda$. This condition puts a constraint on the experimental realization of the subtraction beam splitter interaction transmissivity $t_s^2$: it should be not less than $\sim50\%$. On the other hand, for moderate values of $t_s$, the critical $m_c$ properties of the protocol does not differ much  with respect to changes in the time-bandwidth product $\tau$. The difference becomes significant for values of $t_s$ close to one. Note that the quantity $m_c$ additionally contains the memory lifetime property. Namely, $m_c$ does not increase infinitely: its upper bound value $m_c^u$ is of the order of the time-bandwidth product $\tau$ ($m_c^u\sim\tau$).

As an additional measure of the efficiency of the protocol, which incorporates its probabilistic nature, we consider the average entanglement rate $\langle E \rangle$. We define it using the above discussed quantity $m_c$:
\begin{equation}
\langle E \rangle=\sum_{j=1}^{m_c} P^f_{m_A, j} N_f(\rho), \label{12}
\end{equation}
where $P^f_{i, j}$ corresponds to the joint probability of detecting a phonon at $i^\mathrm{th}$ and $j^\mathrm{th}$ attempts (on arms $A$ and $B$ respectively), with a consequent success probability of a vacuum detection in the mashing step. The quantity $N_f(\rho)$ stands for the negativity of the final state at the output of the distillery. Here we put an additional constraint on the phonon subtraction on arm $A$: we assume that it occurs at the first step, i.e., $m_A=1$. Additionally, we truncate the sum in Eq.~(\ref{12}) to incorporate the fact that the protocol is restarted if one does not succeed to malt the initial two-mode squeezed state within $m_c$ attempts.

\begin{figure}[h!]
\begin{center}
\includegraphics[width=9cm]{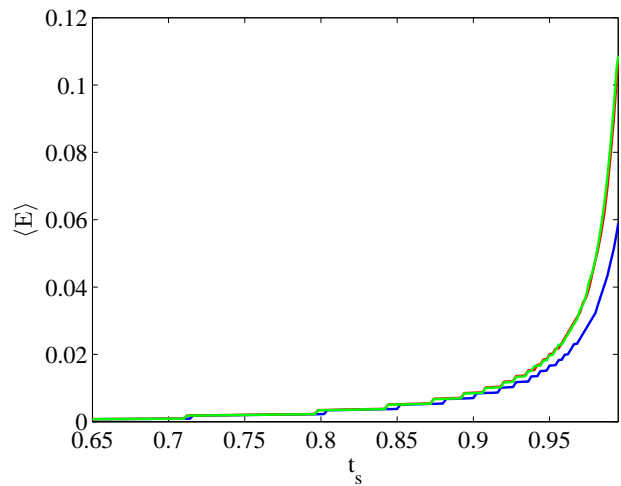}
\caption {The average entanglement rate $\langle E \rangle$ defined in Eq.~(\ref{12}), versus transmissivity $t_s$, for $\lambda=0.1$ and $m_A=1$. Here  $\tau=1/(1-t^2)=100$ (blue curve), 1000 (red curve) and 10000 (green curve). \label{ave}}
\end{center}
\end{figure}

Figure~\ref{ave} shows that one can increase the average entanglement rate by increasing the subtraction beam splitter interaction transmissivity $t_s$. This is an obvious consequence of the fact that higher values of $t_s$ allow a larger number of subtraction attempts, each of which contributes to the above defined average entanglement. Note, however, that the limit $t_s\rightarrow 1$ reduces the probability of a successful malting (time required to finish the malting step tends to infinity), which should result in a decrease of the average entanglement for values of $t_s$ sufficiently close to one. One does not find this effect in the above figure, because of the assumption of a perfect mashing step. The additional entanglement gain due to the increase of $t_s$ compensates the probability decrease.

\section{Conclusion}\label{conc}

We have investigated the compact continuous-variable entanglement distillation protocol in the presence of dissipation. We applied the beam splitter model for losses to track the dynamics of a two-mode squeezed state that is distributed over a pair of quantum memories. Our study showed that dissipation can strongly affect the properties of the original distillation protocol. In particular, phonon retrieval from the matter modes of the quantum memories may not result in entanglement increase. For the analysis of this effect we considered the critical number $m_c$ of phonon subtraction attempts, within which one still gets a gain in entanglement. While studying its properties we assumed that successful phonon retrieval on one of the modes occurs at the first attempt. Since the phonon subtraction probability decreases exponentially with the number of subtraction attempts, this option guarantees the highest success probability and the lowest amount of decoherence effects. We showed that $m_c$ is strictly dependent on the subtraction beam splitter interaction transmissivity $t_s^2$. More precisely, there exists a threshold value of $t_s$, below which even a success at the first malting attempt does not lead to entanglement increase. This puts a constraint on the experimental realization of the subtraction beam splitter interaction transmissivity. Namely, it should be higher than $\sim50\%$ for a broad range of the protocol parameters. Additionally, we found that the value of $m_c$ can be increased by maximizing $t_s$. Note that the transmissivity $t_s^2$ can be considered as a free parameter, subject to manipulation in the experiment. On the other hand, taking into account the probabilistic nature of the protocol, we considered the average entanglement rate, defined as the average negativity within $m_c$ subtraction attempts. Our calculations showed that the strategy aimed to increase this quantity also consists in maximizing $t_s$. However, for evaluating the exact upper bound of $t_s$, one has to take into account the dissipation effects in the mashing step, which we will address in our further works.

We note that the assumption of a perfect mashing step, that we adopted here to simplify the analysis, is justified by the fact that the possible entanglement increase is mainly provided by malting, and so the effects of loss on this process provide insight into the robustness of the protocol. On the other hand, even with this simplification, the present study is relevant to the next generation of experiments attempting to implement this protocol, because initial demonstrations would be focused on the technically simpler malting step. Note that mashing (or its equivalent), has not yet been attempted in any entanglement distillation experiment so far, because of its complexity.

Finally, we note that the protocol presented here could be implemented using emerging technologies. Several groups are developing high efficiency quantum memories \cite{gem, afc, yu_eit} and memories with very long storage times \cite{heinze_eit}. Large values for the time-bandwidth product $\tau$ have already been demonstrated, exceeding several thousands \cite{bao_ring_cavity, mem1}, and in some cases achieving values as high as $\tau \sim 10^8$ \cite{kuzmich_eit}. The formal analogy between the interaction in a Raman memory and a beam splitter, upon which this protocol is based, has also been demonstrated experimentally \cite{campbell_beam_splitter, reim_beam_splitter}. Here it was found that a wide range of beam splitting ratios can be readily achieved, so that the constraints derived above can be readily satisfied. Concerning the detectors required for this protocol, we note that fast photodetectors with timing jitter on the order of tens of picoseconds are now a well established technology \cite{cova_spad}. In the current protocol, repeated detections are not required (each detector only needs to fire once to implement photon subtraction, for instance) so detector dead-time plays no role in limiting the clock speed for the protocol. Finally, high efficiency photodetectors based on extremely sensitive phase transitions in superconducting materials are set to enable vacuum detection and photon number resolution \cite{detect, trans}. We therefore expect that the entanglement distillery we have modelled here, including realistic losses, will be feasible in the near term.

\section*{Acknowledgements}
This work has been supported by the Royal Society, EU IP QESSENCE (248095), EPSRC (EP/J000051/1 and EP/H03031X/1), AFOSR EOARD (FA8655-09-1-3020), FASTQUAST ITN Program of the 7$th$ FP and EU IP SIQS (600645). L.C. gratefully acknowledges the support from the Conseil R\'{e}gional de Bourgogne.

\end{document}